\documentclass{jjap3}
\usepackage{bm}

\newcommand{\rbk}[1]{\left( #1 \right)}
\newcommand{\brc}[1]{\left\{ #1 \right\} }
\newcommand{\sbk}[1]{\left[ #1 \right]}

\newcommand{\abs}[1]{\left| #1 \right|}

\newcommand{\dif}[2]{\frac{d #1}{d #2}}

\newcommand{\pdif}[2]{\frac{\partial #1}{\partial #2}}

\newcommand{\arrcase}[1]{\left\{ \begin{array}{ll} #1 \end{array}\right.}
\newcommand{\vc}[1]{\bm{#1}}

\title{Formation and Classification of Amorphous Carbon	 by Molecular Dynamics Simulation}

\author{Atsushi M. {Ito}$^{1}$\thanks{E-mail address: ito.atsushi@nifs.ac.jp},
 Arimichi {Takayama}$^{1}$, Seiki {Saito}$^{2}$, and Hiroaki {Nakamura}$^{1,2}$}

\inst{$^{1}$Department of Helical Plasma Research, National Institute for Fusion Science, 322-6 Oroshicho, Toki, Gifu 509-5292, Japan\\
$^{2}$Department of Energy Engineering and Science, Nagoya University, Furocho, Chikusa-ku, Nagoya 464-8602, Japan}

\abst{%
By using molecular dynamics simulation, formation mechanisms of amorphous carbon in particular sp${}^3$ rich structure was researched.
The problem that reactive empirical bond order potential cannot represent amorphous carbon properly was cleared in the transition process from graphite to diamond by high pressure and the deposition process of amorphous carbon thin films.
Moreover, the new potential model which is based on electron distribution simplified as a point charge was developed by using downfolding method.
As a result, the molecular dynamics simulation with the new potential could demonstrate the transition from graphite to diamond at the pressure of 15 GPa corresponding to experiment and the deposition of sp${}^3$ rich amorphous carbon.
}

\kword{carbon, amorphous carbon, graphite, diamond, phase transition, diamond like carbon, molecular dynamics, potential}

\begin{document}
\maketitle
\thispagestyle{empty}

\section{Introduction}

There is a lot of amorphous material in the world. In the past, almost all of them had been classified into `amorphous' just because they do not have crystalline structures.
However, materials can be treated in nano-meter scale currently, and then the fact that the amorphous materials with different structures and properties elements in spite of same constituent atomic elements exist is emerged.
The reason for the structural diversity of amorphous carbon materials is that an sp${}^3$ structure which is a metastable state is as much stable as an sp${}^2$ structure in carbon. 
From this reason, one of important parameter to classify amorphous carbon materials is the sp${}^3$/sp${}^2$ ratio, which is the ratio of the number of sp${}^3$ carbon atoms to that of sp${}^2$ carbon atoms. 
Actually, in also the most well-known classification of amorphous carbon materials proposed by W. Jacob, W. M\={o}ller and Robertson\cite{Jacob, Robertson1}, the sp${}^3$/sp${}^2$ ratio was selected as a classification parameter with hydrogen content rate. 
However, the amorphous carbon materials which are same in sp${}^3$/sp${}^2$ ratio and hydrogen and but differ in physical property such as strength exist. 
The classification by these two parameters is, therefore, not enough to understand amorphous carbon materials.
In other words, if more detail classification is obtained, it can lead to the discovery of the new amorphous carbon materials which have physical property functionality such as friction property, electric conduction property, semiconductor property, and so on.
Aiming to establish new classification of amorphous carbon materials, we have investigated by using molecular dynamics (MD) simulation.
MD simulation is powerful method because it can demonstrate amorphous structures in atomic scale directly. 
Actually, our previous research cleared that the amorphous carbon materials created by deposition process or annealing process differ in the orientation of covalent bonds although they are similar in density, sp${}^3$/sp${}^2$ ratio and radial distribution function\cite{Saito_ISPlasma2011}.
It follows from this that to clear the difference of amorphous carbon materials, their formation processes should be understood simultaneously.

The purpose of this study is to investigate the amorphous carbon materials of sp${}^3$ rich structures. 
In general, almost of all carbon materials have sp${}^2$ rich structures and diamond, which is typical material of an sp${}^3$ rich structure, is generated in high pressure environment.
However, diamond like carbons (DLCs) are created by deposition process in the condition of low pressure and then they have sp${}^3$ rich structures. It is not clear how DLCs can have a lot of sp${}^3$ carbon atoms in amorphous structure.

In the previous work, we investigated amorphous carbon deposition by using MD simulation with the second generation reactive empirical bond order (REBO2002) potential\cite{Brenner2002,Ito_jpsj}. 
As a result, relationship between deposition ratio and H/C ratio of incident atoms onto a surface, and then the relationship agrees with experimental result\cite{Ito_ISPlasma2010}. 
After that, using similar MD simulation we have tried to clear the difference between formation process of sp${}^2$ and sp${}^3$ rich amorphous carbon deposits. 
Simultaneously, we have tackled graphite to diamond transition under high pressure because the transition is regarded as a structure change from a pure sp${}^2$ structure of graphite to a pure sp${}^3$ structure of diamond via sp${}^2$ rich and sp${}^3$ rich amorphous carbons structures. 
It is important to study from various viewpoints to elucidate the complicate mechanism of amorphous carbon. 
However, we encountered the obstacles that sp${}^3$ rich amorphous carbon cannot created in MD simulation and graphite cannot change into diamond by more high pressure than 100 GPa, where experimental transition pressure was reported by about 15 GPa\cite{ExpG2D15GPa}. 
We consider the reason of this problems is that the empirical function forms in the REBO2002 potential have trouble representing potential energy for amorphous structures.

Here, we note the empirical function forms in the REBO2002 potential. 
In general, almost all model of potential in MD of real atomic system, in which dominant interaction is covalent bonding especially, is classified into `bond order (BO)' potential. 
BO potential is composed of the functions of the length of covalent bond, the bond angle which is the angle between adjacent two covalent bonds, the dihedral angle which is the angle between adjacent two planes formed by continuous three covalent bonds, and so on. 
As a extension for BO potential to treat new connecting and cutting of covalent bonds, Brenner et al. added the term of functions of the number of surrounding atoms within a cutoff length. 
This surrounding term represents the energy difference among molecular structures empirically, and then the potential energy of a reaction intermediate molecule was given by interpolation of energies of molecules before and after reaction, which is called reactive empirical bond order (REBO) potential\cite{Brenner90}. 
The REBO potential is most often employed in the study of carbon nano-materials since the discovery of carbon nanotube and fullerene. 
Moreover, the second generation model of REBO potential (REBO2002), which had been improved over a period of twelve years by Brenner et al.\cite{Brenner2002}, can calculate not only reasonable energies for almost all carbon and hydro-carbon molecules.

Thus, the REBO2002 potential well represents stable molecule and lattice structures in potential energy. 
However, meta-stable structures such as amorphous structures are hardly considered in the REBO2002 potential and general BO potentials. 
Therefore, in the present work, we propose new potential model for amorphous carbon material. 
The new potential model is based on consideration of electron distribution like molecular orbital, and is not BO and REBO potentials. 
Consequently, in the MD simulation with the new potential model, the creation of sp${}^3$ rich amorphous carbon in a deposition process and the graphite to diamond transition under pressure at 15 GPa could succeed.

\section{Electron Order Potential}

\subsection{Potential function form}
\label{ss:2.1}
As a function form of the new potential, BO and REBO type functions are not employed. 
Then, from natural viewpoint, total potential energy $U_\mathrm{tot}$ is composed of nucleus-nucleus repulsive energy $U_\mathrm{nn}$, nucleus-electron attractive energy $U_\mathrm{ne}$, and electron-electron repulsive energy $U_\mathrm{ee}$:
\begin{eqnarray}
	U_\mathrm{tot} = U_\mathrm{nn} + U_\mathrm{ne} + U_\mathrm{ee}
\label{eq1}
\end{eqnarray}

The nucleus-nucleus repulsive energy $U_\mathrm{nn}$ has a function form of screened Coulomb interaction given by 
\begin{eqnarray}
	U_\mathrm{nn} = \frac{1}{2}\sum_{i,j \neq i} K_\mathrm{nn} \frac{Q_i Q_j}{r_{ij}} \exp \rbk{ -a_\mathrm{nn} r_{ij}} f_\mathrm{nn}(r_{ij}),
\label{eq2}
\end{eqnarray}
where $\sum_{i,j \neq i}$ means summation in terms of pair of $i$-th and $j$-th nuclei, $Q_i$ and $Q_j$ are core charge of nuclei which are screened by core electrons,
$r_{ij} = \abs{\vc{r}_i - \vc{r}_j}$ is a distance between the $i$-th and $j$-th nuclei, and $f_\mathrm{nn}(r_{ij})$ is cutoff function.
The detail of cutoff functions is described later.

To construct the nucleus-electron attractive energy $U_\mathrm{ne}$ and electron-electron repulsive energy $U_\mathrm{ee}$, electron distribution like molecular orbitals is considered.
Of course, the electron distribution is simplified as a point charge located on the position of atom (nuclei) $\vc{r}_i$ and the center position between pair atoms $\bar{\vc{r}}_{ij} = (\vc{r}_i + \vc{r}_j)/2$.
When the $i$-th and $j$-th atoms are close, electric charge transfer from the position of the $i$-th atom to the center position between $i$-th and $j$-th atoms is defined by
\begin{eqnarray}
	q_{ij}(r_{ij}) = f_\mathrm{q}(\frac{r_{ij} - r_\mathrm{es}}{r_\mathrm{ed}})   \label{eq3}\\
	f_\mathrm{q}(r) = \arrcase{ 0 & \textrm{if}\indent r > 1, \\
	(- 6 r^2 + 15 r - 10) r^3 + 1  & \textrm{if}\indent 1 \geq r > 0, \\
	0 & \textrm{else,} \\}			\label{eq4}
\end{eqnarray}
where single electric charge is normalized as 1.
In this definition, if $r_{ij} > r_\mathrm{es} + r_\mathrm{ed}$, $q_{ij}(r_{ij}) = 0$. That is, electron is unpaired electron completely.
Therefore the electric charge of unpaired electron located on the position of the $i$-th atom is 
\begin{eqnarray}
	\bar{q}_{i} = q^\mathrm{v}_i - \sum_{j\neq i}q_{ij}(r_{ij})
\label{eq5}
\end{eqnarray}
where $q^\mathrm{v}_i$ is number of valence electrons depending on the atomic element of the $i$-th atom.

The nucleus-electron attractive energy $U_\mathrm{ne}$ is divided into four terms as follows:
\begin{eqnarray}
	U_\mathrm{ne} = U_\mathrm{ne}^\mathrm{s} + U_\mathrm{ne}^\mathrm{s2}  + U_\mathrm{ne}^\mathrm{u} + U_\mathrm{ne}^\mathrm{u2}
\label{eq6}
\end{eqnarray}
The term $U_\mathrm{ne}^\mathrm{s}$ is the attractive energy between nucleus and the nearest shared electron pair defined by
\begin{eqnarray}
	U_\mathrm{ne}^\mathrm{s} = - \sum_{i,j\neq i} K_\mathrm{ne}^\mathrm{s} \exp \rbk{ -a_\mathrm{ne}^\mathrm{s} \abs{\bar{\vc{r}}_{ij} - \vc{r}_i}} Q_i \sbk{q_{ij}(r_{ij}) + q_{ji}(r_{ji})}.
\label{eq7}
\end{eqnarray}
The term $U_\mathrm{ne}^\mathrm{s2}$ is the attractive energy between nucleus and the second nearest shared electron pair defined by
\begin{eqnarray}
	U_\mathrm{ne}^\mathrm{s} = - \sum_{i,j\neq i, k\neq (i,j)} K_\mathrm{ne}^\mathrm{s2} \exp \rbk{ -a_\mathrm{ne}^\mathrm{s2} \abs{\bar{\vc{r}}_{ik} - \vc{r}_j}} Q_j \sbk{q_{ik}(r_{ik}) + q_{ki}(r_{ki})} f_\mathrm{ne}^\mathrm{s2}(r_{ij}),
\label{eq8}
\end{eqnarray}
where $f_\mathrm{ne}^\mathrm{s2}(r_{ij})$ is cutoff function.
The term $U_\mathrm{ne}^\mathrm{u}$ is the attractive energy between nucleus and the nearest unpaired electron defined by
\begin{eqnarray}
	U_\mathrm{ne}^\mathrm{u} = - \sum_{i} K_\mathrm{ne}^\mathrm{u} Q_i \sbk{q^\mathrm{v}_i - \sum_{j\neq i}q_{ij}(r_{ij})},
\label{eq9}
\end{eqnarray}
where the electric charge of unpaired electron is given by eq. (\ref{eq5}). The nearest unpaired electron is located on same position to nucleus.
The term $U_\mathrm{ne}^\mathrm{u}$ is the attractive energy between nucleus and the second nearest unpaired electron defined by
\begin{eqnarray}
	U_\mathrm{ne}^\mathrm{u2} = - \sum_{i,j\neq i} K_\mathrm{ne}^\mathrm{u2} \exp \rbk{ -a_\mathrm{ne}^\mathrm{u2} r_{ij} } Q_j \sbk{q^\mathrm{v}_i - \sum_{k\neq i}q_{ik}(r_{ik})} f_\mathrm{ne}^\mathrm{u2}(r_{ij}),
\label{eq10}
\end{eqnarray}
where $f_\mathrm{ne}^\mathrm{s2}(r_{ij})$ is cutoff function. Because the second nearest unpaired electron is located on same position to the second nearest nucleus, the distance between nucleus and the second nearest unpaired electron is same to that between pair of nuclei.

The electron-electron repulsive energy $U_\mathrm{ee}$ is modeled by using the electric charge of shared electron pair of eq. (\ref{eq3}) and that of unpaired electron of eq. (\ref{eq5}) similarly to $U_\mathrm{ne}$.
The electron-electron repulsive energy $U_\mathrm{ee}$ is composed of the following three terms:
\begin{eqnarray}
	U_\mathrm{ee} = U_\mathrm{ee}^\mathrm{p} + U_\mathrm{ee}^\mathrm{ss}  + U_\mathrm{ee}^\mathrm{su}
\label{eq11}
\end{eqnarray}
The first term $U_\mathrm{ee}^\mathrm{p}$ is repulsive energy between two electrons of shared pair defined by
\begin{eqnarray}
	U_\mathrm{ee}^\mathrm{p} = \frac{1}{2}\sum_{i,j\neq i} K_\mathrm{ee}^\mathrm{p} q_{ij}(r_{ij})q_{ji}(r_{ji}),
\label{eq12}
\end{eqnarray}
The second term $U_\mathrm{ee}^\mathrm{ss}$ is repulsive energy between adjacent two shared electron pairs defined by
\begin{eqnarray}
	U_\mathrm{ee}^\mathrm{ss} = \sum_{i,j\neq i, k > j} K_\mathrm{ee}^\mathrm{ss} \exp \rbk{ -a_\mathrm{ee}^\mathrm{ss} \abs{\bar{\vc{r}}_{ij} - \bar{\vc{r}}_{ik}} } \sbk{q_{ij}(r_{ij}) + q_{ji}(r_{ji})} \sbk{q_{ik}(r_{ik}) + q_{ki}(r_{ki})}.
\label{eq13}
\end{eqnarray}
The third term $U_\mathrm{ee}^\mathrm{su}$ is repulsive energy between shared electron pair and the nearest unpaired electron defined by
\begin{eqnarray}
	U_\mathrm{ee}^\mathrm{su} = \sum_{i,j\neq i} K_\mathrm{ee}^\mathrm{su} \exp \rbk{ -a_\mathrm{ee}^\mathrm{su} \abs{\bar{\vc{r}}_{ij} - \vc{r}_i}} \sbk{q_{ij}(r_{ij}) + q_{ji}(r_{ji})} \sbk{q^\mathrm{v}_i - \sum_{k\neq i}q_{ik}(r_{ik})}.
\label{eq14}
\end{eqnarray}

There are three cutoff function $f_\mathrm{nn}$, $f_\mathrm{ne}^\mathrm{s2}$, $f_\mathrm{ne}^\mathrm{u2}$ in the above definitions. In the present work, all cutoff functions are unified by using eq. (\ref{eq4}) as
\begin{eqnarray}
	f_\mathrm{nn}(r) = f_\mathrm{ne}^\mathrm{s2}(r) = f_\mathrm{ne}^\mathrm{u2}(r) =  f_\mathrm{q}(\frac{r - r_\mathrm{es}}{r_\mathrm{ed}})
\label{eq15}
\end{eqnarray}

Except for the nucleus-nucleus repulsive energy $U_\mathrm{nn}$ in eq. (\ref{eq2}), all terms have no Coulomb potential form.
This reason is that $U_\mathrm{ne}$ and $U_\mathrm{ee}$ are regarded as interaction energy of not so much classical electron particle but electric charge distribution due to quantum electron orbitals in the present model.
Although the potential model in the present work is composed of only the above terms, higher quality potential model can be created by adopting nucleus-electron attractive terms and electron-electron repulsive terms between more distant positions.

\subsection{Downfolding method}

Ideal potential model should be able to calculate proper inter-atomic interaction energy in arbitrary atomic geometry. 
In other word, the potential model which can calculate inter-atomic interaction energy in many kinds of atomic geometries with accuracy is regarded as good model.
Currently, development of the quantum chemistry which is represented by density functional theory (DFT) are made it possible to calculate the inter-atomic interaction energy with a high degree of accuracy. 
Then, by converging the parameters in potential function so as to reduce the difference between inter-atomic interaction energies calculated by potential model and DFT in many atomic geometries, we can obtain good potential model. 
Yoshimoto proposed the downfolding method to develop potential model from these point of view\cite{Yoshimoto_Downfolding}. 
In the present work, the parameters in the function form given in \S \ref{ss:2.1} are optimized by the downfolding method as follows.

Potential is, of course, a function of atomic geometry $\brc{\vc{r}}=\brc{\vc{r}_1, \vc{r}_2, \cdots}$. 
If parameters in potential function are regarded as variables, the potential is also a function of the parameters. 
The parameters in potential function is described as $\brc {a} = \brc{a_1, a_2, \cdots}$, and the potential function is described as $U(\brc{\vc{r}}, \brc{a})$.
The variance of difference between inter-atomic interaction energy calculated by using potential model and DFT in terms of $N$ kinds of sample atomic geometries is given by 
\begin{eqnarray}
	\Phi = \frac{1}{N}\sum_i \sbk{ U(\brc{\vc{r}}_i, \brc{a}) - E_\mathrm{ref}(\brc{\vc{r}}_i) }^2
\label{eq16}
\end{eqnarray}
where $\brc{\vc{r}}_i$ means the i-th sample atomic geometry and the reference energy $E_\mathrm{ref}(\brc{\vc{r}}_i)$ is the inter-atomic interaction energy calculated by using DFT. 
As so to reduce $\Phi$, the parameters $\brc{a}$ is optimized.

One way to optimize the parameters $\brc{a}$ is to solve the following evolution equation on a virtual time $t$:
\begin{eqnarray}
	\dif{a_k^t}{t} = - c \pdif{\Phi}{a_k^t} = - \frac{2c}{N} \sum_i \pdif{U(\brc{\vc{r}}_i, \brc{a^t})}{a_k^t} \sbk{ U(\brc{\vc{r}}_i, \brc{a^t}) - E_\mathrm{ref}(\brc{\vc{r}}_i) } ,
\label{eq17}
\end{eqnarray}
where the parameters $\brc{a^t} = \brc{a_1^t, a_2^t, cdots}$ indicate the parameters at the virtual time $t$ and the coefficient $c$ is weight for time evolution. 
The parameters $\brc{a^t} $ when $t \rightarrow \infty$ is optimal parameters.
In the numerical solvent, this evolution equation is replaced with the difference equation given by 
\begin{eqnarray}
	a_k^{t+1} = a_k^t - \frac{2c \Delta t}{N} \sum_i \pdif{U(\brc{\vc{r}}_i, \brc{a^t})}{a_k^t} \sbk{ U(\brc{\vc{r}}_i, \brc{a^t}) - E_\mathrm{ref}(\brc{\vc{r}}_i) }.
\label{eq18}
\end{eqnarray}

In actual operation, how to prepare the sample atomic geometries is important.
Yoshimoto proposed that the sample atomic geometries are selected from animation snapshot in preliminarily performed MD simulation with the potential function $U(\brc{\vc{r}}, \brc{a^0})$ which employs temporary parameters $\brc{a^0}$. 
For example, if the preliminary MD simulation is executed in canonical ensemble scheme, the sample atomic geometries are generated according to canonical distribution. 
Yoshimoto adopted multi-canonical ensemble scheme as a preliminary MD simulation. 
Moreover, the iteration of the above sequence of the downfolding method, replacing temporary parameters $\brc{a^0}$ with the optimal parameters given by the previous sequence, is an effective technique to obtain more optimal parameters. 
This iteration is also effective to prevent the parameters from being trapped in local minimum.

In the present work, the optimized parameters $\brc{a}$ in the above sequence are corresponding to fifteen parameters $\{ Q$, $K_\mathrm{nn}$, $a_\mathrm{nn}$, $K_\mathrm{ne}^\mathrm{s}$, $a_\mathrm{ne}^\mathrm{s}$, $K_\mathrm{ne}^\mathrm{s2}$, $a_\mathrm{ne}^\mathrm{s2}$, $K_\mathrm{ne}^\mathrm{u}$, $K_\mathrm{ne}^\mathrm{u2}$, $a_\mathrm{ne}^\mathrm{u2}$, $K_\mathrm{ee}^\mathrm{p}$, $K_\mathrm{ee}^\mathrm{ss}$, $a_\mathrm{ee}^\mathrm{ss}$, $K_\mathrm{ee}^\mathrm{su}$, $a_\mathrm{ee}^\mathrm{su}$  $\}$ in the function form defined in \S \ref{ss:2.1}.
Sample atomic geometries were created by the three kinds of preliminary MD simulation with REBO2002 potential. 
One third of sample atomic geometries were selected from graphite structures vibrating at high temperature, the next one third of sample atomic geometries were selected from diamond structure similarly, and the other one third of sample atomic geometries were selected from amorphous structures made by pressing graphite in the preliminary MD simulation.
The reference energy $E_\mathrm{ref}(\brc{\vc{r}}_i)$ for the sample atomic geometries were calculated with 'Open source package for Material eXplorer' (OpenMX)\cite{OpenMX}, which is numerical software based on DFT. 
The generalized gradient approximation (GGA) with Perdew-Burke-Ernzerhof (PBE) functional\cite{GGA-PBE} was employed as exchange-correlation potential. 
Norm conserving pseudo-potentials\cite{VPS} were used for approximation for core electrons and nuclei. 
Electron orbital was represented by numerical orbital with pseudo-atomic localized basis functions. 

As a result, the optimal parameters were obtained, where the iteration of the downfolding method was not performed. 
The optimal parameters and fixed parameters are shown in Table \ref{tb:param}.

\section{Application}
\subsection{Transition from graphite to diamond} \label{ss:3.1}

The first results by the new potential was that the transition from graphite to diamond could be well imitated by MD simulation. 
In this MD simulation, it was investigated that the lattice structure changes from a graphite structure into a diamond structure as the pressure of system increases gradually.
The system had a 24 carbon atoms, and initially they construct the graphite structure which was composed of three graphene layers stacked as a 'ABC' form.
The simulation box was under the periodic boundary condition.
The pressure of system was controlled by the Andersen's method[ref], where a setting pressure increased from 3 GPa to 15 GPa in the first 4 x $10^{-12}$s, and then it was kept at 15 GPa during the next 4 x $10^{-12}$s.
Though the size of simulation box was initially 0.492 x 0.426 x 1.004 nm${}^3$, it was changed according to the pressure of system on the basis of the procedure of the Andersen's method.
Different point from the original Andersen's method was that the pressure and the size of simulation box were controlled independently for each $x$, $y$, $z$ directions.
The temperature of the system was also controlled by the Langevin thermostat method, in which a setting temperature was 1000 K and a friction coefficient is 1.0 x $10^14$ s$^{-1}$.
The time step of the MD simulation was 4.0 x $10^{-17}$s. 
After 2 x $10^5$ steps, the MD simulation finished.

Figure \ref{fig:g2dpressure} shows the change of pressures in the MD simulation. 
From this figure, it was confirmed that the pressure acting actually on carbon atoms well follows the setting pressure of the Andersen' s method. 
In this simulation, we could demonstrate the transition from graphite to diamond at the pressure of 15 GPa, which is agreement with experimental report\cite{ExpG2D15GPa}, as shown in Fig. \ref{fig:g2danime}. 
Quantitative representation of the transition from graphite to diamond was given by the change of ratio of the numbers of sp${}^2$ and sp${}^3$ carbon atoms to that of all carbon atoms as Fig. \ref{fig:g2dspratio}, which are here called sp${}^2$ and sp${}^3$ ratios, simply. 
As the pressure increased, the sp${}^2$ ratio decreased and the sp${}^3$ ratio increased. 
When the pressure reached 15 GPa, the sp${}^3$ ratio also achieve to become 1.0. 
That is, carbon atoms became a perfect sp${}^3$ material. 
From the simulation snapshot shown in Fig. \ref{fig:g2danime}(last), it was confirmed that when the sp$^3$ ratio was 1.0, the carbon atoms certainly constructed a diamond structure, not an amorphous structure.

\subsection{Deposition of amorphous carbon}

Using the present new potential model, we tried to simulate the deposition generating an sp${}^3$ rich amorphous carbon. 
The manner of the MD simulation of deposition was similar to our previous work \cite{Ito_ISPlasma2010}. 
The substrate was prepared by piling the amorphous carbon blocks which was created by the MD simulation for deposition on the REBO2002 potential \cite{Saito_ISPlasma2011}. 
The sp${}^3$ ratio of the substrate was about 20 percent. 
The size of the surface of the substrate was 2.019 x 2.186 nm${}^2$ and then the substrate was put on the simulation box which follows periodic boundary condition in the parallel direction to the surface of the substrate. 
The carbon atoms initially putted on simulation box as the substrate were connected to the Langevin thermostat with a friction coefficient of 1.0 x $10^14$ s$^{-1}$ to control temperature, where the motion of carbon atoms located in the range of up to 0.12 nm from the bottom position was fixed during simulation. 
In one simulation, 1000 carbon atoms as a source were continuously injected from 3.4 nm above the initial surface of the substrate every 1.0 x $10^{12}$ s. 
Injection position in parallel direction to the surface of the substrate was determined randomly under uniform distribution. 
These MD simulations were executed varying the injection energy from 1 eV to 200 eV.

As a result of the MD simulation using the present new potential, we could obtain the sp${}^3$ rich amorphous carbons as deposits. 
Figure \ref{fig:cvdspratio} shows sp${}^2$ and sp${}^3$ ratios of only the injected carbon atoms staying in the deposit, except for that constructing initial substrate and that reflected from a surface. 
From this figure, it was confirmed that the sp${}^3$ ratio in the case of the present new potential was grater than that of the REBO2002 potential for whole injection energies.

The penetration depth of the injected carbon atoms was also investigated as shown in Fig. \ref{fig:cvdtddepth}. 
Here, the penetration depth was defined as the mean of the distance between the surface position at the moment of each injection and the final position of each injected carbon atoms. 
In the case using the present new potential, the penetration depth increased as the injection energy increased higher than 20 eV, while it was independent of the injection energy for lower range. 
On the other hand, in the case using the REBO2002 potential, the injected carbon atoms hardly penetrated from the surface. 
The penetration depth in case of the injection energy of 200 eV is even corresponding to three times of thickness of diamond and graphite monolayer.

Deposition rate was also estimated by the time evolution of the thickness of deposit as Fig. \ref{fig:cvdthickness}. 
Here, the thickness was defined by the difference of the surface position from the minimum position while simulation. 
The reason why the thickness decreased for a while from start of simulation is by the contraction of substrate. 
The substrate composed of stacked amorphous block, which was created by using the REBO2002 potential, had an sp${}^2$ rich structure and is unstable state for the present new potential, the thickness decreases. 
The simulation in the case of high energy injection used many amorphous blocks to prepare the substrate, and the decreases of thickness was larger. 
After the contraction of substrate, the thickness increased linearly. 
By this linear growth, the deposition rate could be evaluated as the gradient of the increase of the thickness. 
It is understand that the deposition rate was independent of the injection energy of less than 100 eV because sputtering hardly occurred. 
The case of the incident energy of 100 eV or more, we should calculate more long time.

\section{Discussion}

In the present paper, we proposed the new potential model. 
In fact, before developing the new potential model, we had tackled the modification of the REBO2002 potential to represent the sp${}^3$ rich amorphous carbon. 
Especially, in terms of the three body terms depending on bond angles, many function forms were tried and parameters were optimized by using the downfolding method every time. 
However, we could not obtain well modified potential to achieve the transition from graphite to diamond due to high pressure. 
Compromising optimization in parameter, the graphite kept its structure even of pressure over 100 GPa or it changed to highly-oriented structures. 
From this experience, we consider that approximation of interaction of covalent bonding system with the bond angle, dihedral angle and so on, which are historically employed by bond order type potential models, does not give us well representation for amorphous structures, which are metastable structures.
For the REBO potential, it was reported that the elastic constant of diamond is not also properly evaluated\cite{Gao}.

On the other hand, the bond angle and the dihedral angle were not employed as the variables of the function form of the present new potential. 
Although simplified to point charge, electron distribution was considered and then potential function consists of nuclear-nuclear repulsion, nuclear-electron attraction and electron-electron repulsion according to that natural consideration. 
We carried out the downfolding for this function form, and we could smoothly optimize all parameter. 
Consequently, we could amazingly-easily achieved the transition from graphite to diamond by high pressure. 
In Fig. \ref{fig:g2denergy}, potential energies calculated by the present new potential, the REBO2002 potential and DFT were compared in terms of several carbon structures in the MD simulation with the present new potential for the transition from graphite to diamond in \S \ref{ss:3.1}. 
The potential energies by the new potential model are close to those by DFT, while those by REBO2002 were significantly different from those by DFT. 
This comparison implies the advantage of the present new model in treatment of amorphous carbon structures.

Furthermore, the present transition pressure of 15 GPa agreed with experimental report \cite{ExpG2D15GPa}. 
Almost all MD simulation that graphite changes to diamond at comparable pressure to the experiment by the other researchers employed quantum chemical theory to calculate energy and force of atoms\cite{G2D_QM1, G2D_QM2, G2D_QM3, G2D_QM4, G2D_QM5, G2D_QM6, G2D_QM7, G2D_QM8, G2D_QM9}. 
It's interesting to reflect that the present new potential is classified into the category of classical mechanics though it can demonstrate the transition of graphite to diamond at proper pressure.

%
%
%


\begin{table}[tb]
\caption{Compositions of deposits.}
\label{tb:param}
\begin{tabular}{llll}
\hline
optimized parameter & optimal value & fixed parameter & fixed value\\
\hline
$Q$                         & 3.867623     & $r_\mathrm{es}$ & 1.7 \AA \\
$K_\mathrm{nn}$             & 0.946244 eV   & $r_\mathrm{ed}$ & 0.5 \AA \\
$a_\mathrm{nn}$             & 1.106056 \AA   & $q^\mathrm{v} $ & 4 \\
$K_\mathrm{ne}^\mathrm{s}$  & 1.343592 eV   &  &  \\
$a_\mathrm{ne}^\mathrm{s}$  & 0.652446 \AA   &  &  \\
$K_\mathrm{ne}^\mathrm{s2}$ & 0.229646 eV   &  &  \\
$a_\mathrm{ne}^\mathrm{s2}$ & 0.951189 \AA   &  &  \\
$K_\mathrm{ne}^\mathrm{u}$  & 0.712260 eV   &  &  \\
$K_\mathrm{ne}^\mathrm{u2}$ & 1.144283 eV   &  &  \\
$a_\mathrm{ne}^\mathrm{u2}$ & 0.790185 \AA   &  &  \\
$K_\mathrm{ee}^\mathrm{p}$  & 0.439343 eV   &  &  \\
$K_\mathrm{ee}^\mathrm{ss}$ & 1.007662 eV   &  &  \\
$a_\mathrm{ee}^\mathrm{ss}$ & 1.058837 \AA   &  &  \\
$K_\mathrm{ee}^\mathrm{su}$ & 1.307662 eV   &  &  \\
$a_\mathrm{ee}^\mathrm{su}$ & 1.078244 \AA   &  &  \\
\hline
\end{tabular}
\end{table}

\section{Conclusions}

In the present work, we were aware that the region why sp${}^3$ rich amorphous carbon materials are not well generated by MD simulation is that the REBO2002 potential, which is major model, cannot give correct interaction energy in terms of amorphous carbon structures. 
To solve this problem, we developed new potential model, which was modeled according to that natural consideration that interaction among atoms consists of attraction and repulsion between nuclei and electrons. 
Then, electron distribution was simplified as a point charge located on places between atoms and same place to nuclei. 
Consequently, using the present new potential, we could generate sp${}^3$ rich amorphous carbon deposits. 
In addition, we could demonstrate the transition from graphite to diamond at the pressure of 15 GPa, which was corresponding to experimental report \cite{ExpG2D15GPa}. 

To advance the research for amorphous carbon deposition, the treatment of hydrogen atoms is necessary. 
As additive substance, catalyst or impurity, more atomic element is needed. 
Therefore, we should tackle to extend potential model aiming to treat all atomic element.

\section*{Acknowledgments}

We are very grateful to Dr. Y. Yoshimoto for useful discussion and comments for the downfolding method, and to Prof. N. Ohno and Mr. N. Hirata for helpful comments for amorphous carbon deposition. We also would like to Prof. T. Ozaki for offering 'OpenMX' and to Dr. M. Nishio for helpful comment relate to how to use it. Numerical simulations were carried out by the use of the Plasma Simulator at the National Institute for Fusion Science. The present work were supported by KAKENHI, MEXT (No. 19055005 and No. 23710135), NINS Program for Cross-Disciplinary Study, and NIFS Collaboration Research program (NIFS09KNSS002 and NIFS11KNTS008).


\newpage

\noindent \textbf{Figure Captions}\\

\begin{figure}
	\centering
	\resizebox{\linewidth}{!}{\includegraphics{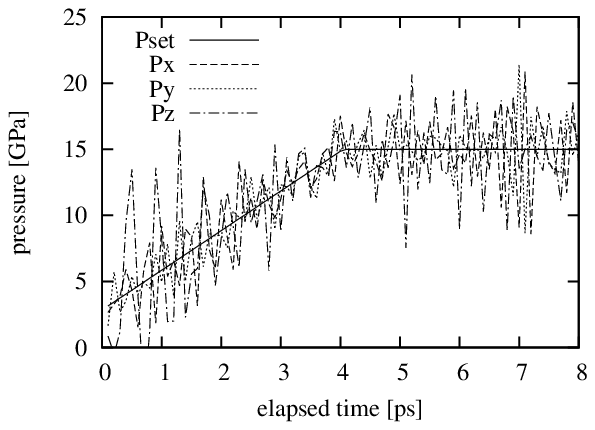}}
	\caption{~}
	\label{fig:g2dpressure}
\end{figure}

\begin{figure}
	\centering
	\resizebox{1.0\linewidth}{!}{\includegraphics{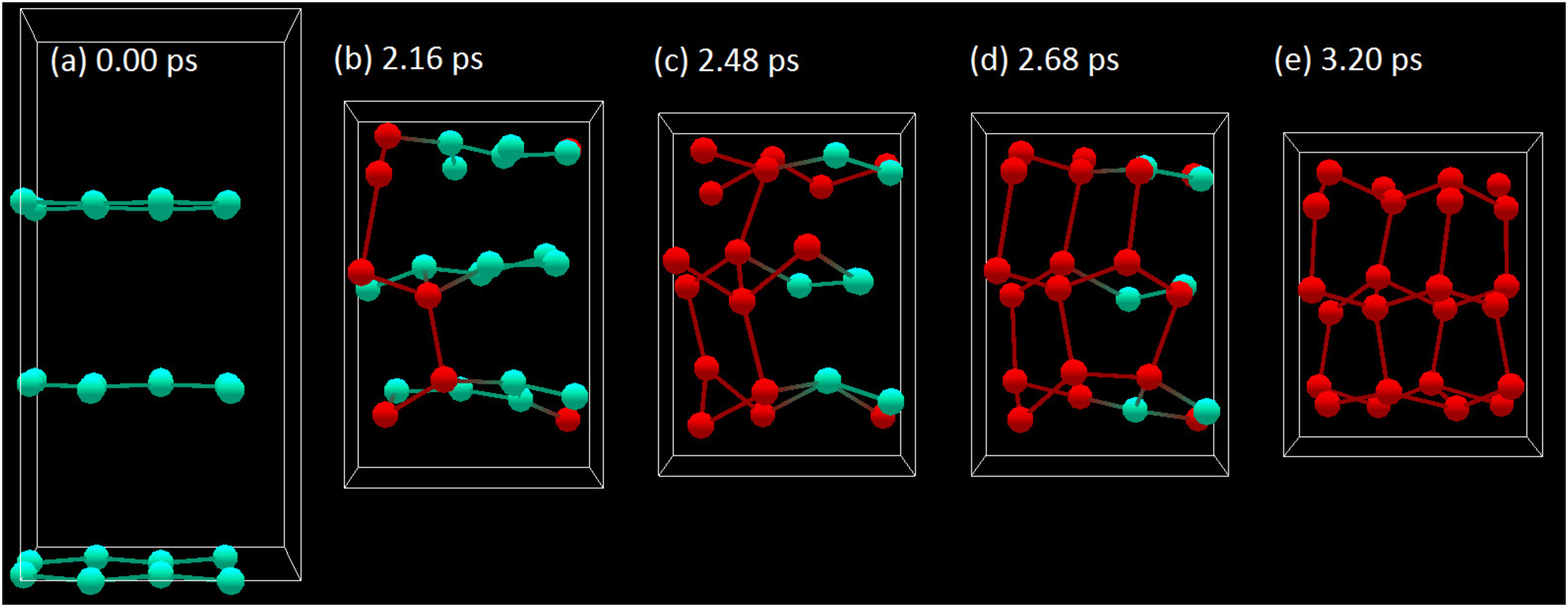}}
	\caption{~}
	\label{fig:g2danime}
\end{figure}

\begin{figure}
	\centering
	\resizebox{\linewidth}{!}{\includegraphics{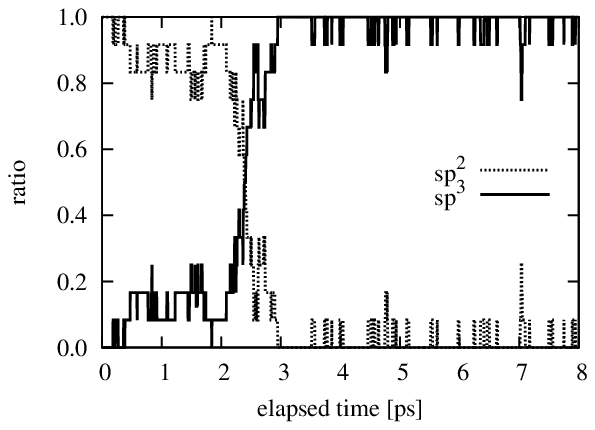}}
	\caption{~}
	\label{fig:g2dspratio}
\end{figure}

\begin{figure}
	\centering
	\resizebox{\linewidth}{!}{\includegraphics{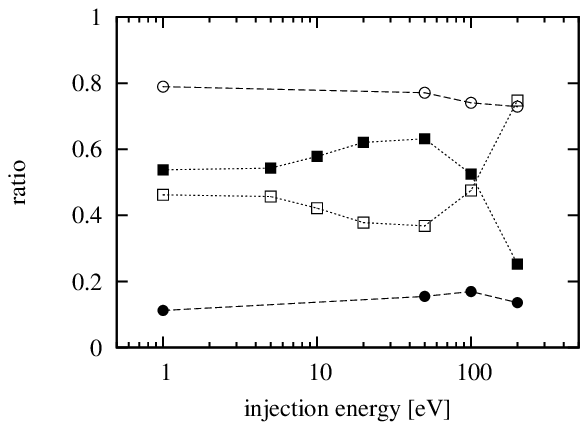}}
	\caption{~}
	\label{fig:cvdspratio}
\end{figure}

\begin{figure}
	\centering
	\resizebox{\linewidth}{!}{\includegraphics{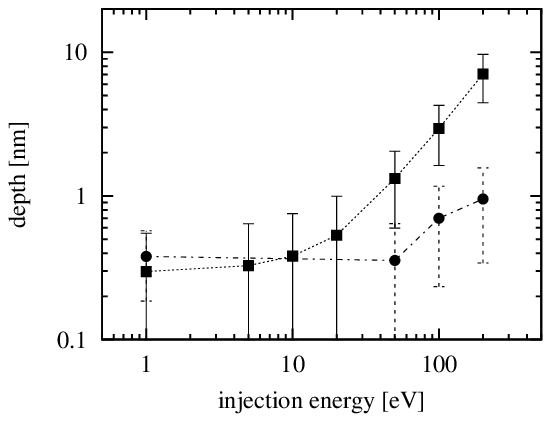}}
	\caption{~}
	\label{fig:cvdtddepth}
\end{figure}

\begin{figure}
	\centering
	\resizebox{\linewidth}{!}{\includegraphics{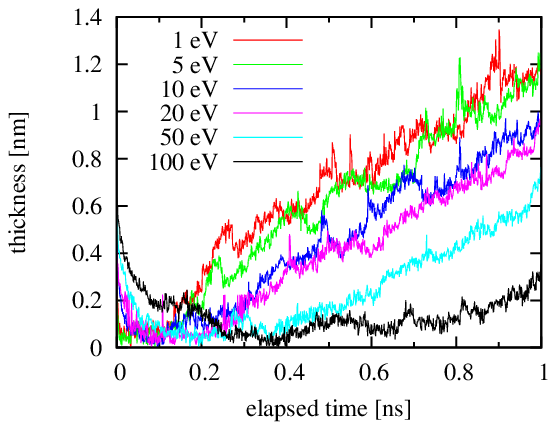}}
	\caption{~}
	\label{fig:cvdthickness}
\end{figure}

\begin{figure}
	\centering
	\resizebox{\linewidth}{!}{\includegraphics{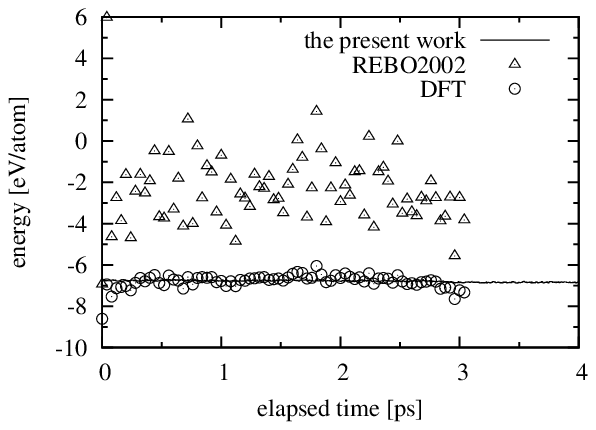}}
	\caption{~}
	\label{fig:g2denergy}
\end{figure}

\begin{list}{}{}

\item[Fig. \ref{fig:g2dpressure}.] {Change of pressure in the MD simulation of transition from graphite to diamond. The solid line indicate the setting pressure in the Andersen's method. The long dashed, short dashed and dashed-dotted lines indicate pressure acting to material for each $x$, $y$, $z$ directions. Their plotted values are mean values for 1.0 x $10^{-13}$ s.}

\item[Fig. \ref{fig:g2danime}.] {(Color online) Animation snapshots in the MD simulation for transition from graphite to diamond at pressure of 15 GPa. The green and red spheres indicate sp${}^2$ and sp${}^3$ carbon atoms. The white lines are the side of simulation box following periodic boundary condition.}

\item[Fig. \ref{fig:g2dspratio}.] {The sp${}^2$ and sp${}^3$ ratios, which is defined as the ratios of the numbers of sp${}^2$ and sp${}^3$ carbon atoms to that of all carbon atoms, as functions of the elapsed time.}

\item[Fig. \ref{fig:cvdspratio}.] {The sp${}^2$ and sp${}^3$ ratio of injected carbon atoms in deposits. The white and black rectangles indicate sp${}^2$ and sp${}^3$ ratios evaluated by the MD simulation with the present new potential, respectively. The white and black spheres indicate these with the REBO2002 potential, respectively.}

\item[Fig. \ref{fig:cvdtddepth}.] {The penetration depth of carbon atoms as a function of injection energy. The penetration depth is mean of that for all injected carbon atoms except for sputtered one. The circle and square points indicate the calculated depths by using REBO2002 potential and the present new potential, respectively.}

\item[Fig. \ref{fig:cvdthickness}.] {(Color online) The thickness of deposit as a function of the elapsed time.}

\item[Fig. \ref{fig:g2denergy}.] {The inter-atomic interaction energies calculated by the present new potential, the REBO2002 potential and DFT using the OpenMX in terms of atomic geometries at the moment of transition process from graphite to diamond in the MD simulation with the present new potential.}

\end{list}

\clearpage
\thispagestyle{empty}

\clearpage
\thispagestyle{empty}

\clearpage
\thispagestyle{empty}

\clearpage
\thispagestyle{empty}

\clearpage
\thispagestyle{empty}

\clearpage
\thispagestyle{empty}

\clearpage
\thispagestyle{empty}

~

\begin{thebibliography}{99} 

\bibitem{Jacob} W. Jacob and W. M\={o}ller, Appl. Phys. Lett. \textbf{63} (1993) 1771.
\bibitem{Robertson1} J. Robertson, Mater. Sci. Eng. R \textbf{37} (2002) 129.

\bibitem{Saito_ISPlasma2011} S. Saito, A. M. Ito, A. Takayama, and H. Nakamura, Jpn. J. Appl. Phys. \textbf{51} (2012) 01AC05.

\bibitem{Brenner2002} D. W. Brenner, O. A. Shenderova, J. A. Harrison, S. J. Stuart, B. Ni, and S. B. Sinnott: J. Phys.: Condens. Matter \textbf{14} (2002) 783.
\bibitem{Ito_jpsj} A. Ito, H. Nakamura, and A. Takayama: J. Phys. Soc. Jpn. \textbf{77} (2008) 114602.
\bibitem{Ito_ISPlasma2010} A. M. Ito, A. Takayama, S. Saito, N. Ohno, S. Kajita, and H. Nakamura, Jpn. J. Appl. Phys. \textbf{50} (2011) 01AB01.

\bibitem{ExpG2D15GPa} F.P. Bundy, J.S. Kasper, J. Chem. Phys. \textbf{46} (1967) 3437.

\bibitem{Brenner90} D. W. Brenner, O. A. Shenderova, J. A. Harrison, S. J. Stuart, B. Ni, and S. B. Sinnott: J. Phys.: Condens. Matter \textbf{14} (2002) 783.

\bibitem{Yoshimoto_Downfolding} Y. Yoshimoto, J. Chem. Phys. \textbf{125} (2006) 184103. 


\bibitem{OpenMX} OpenMX web-site: http://www.openmx-square.org/
\bibitem{GGA-PBE} J. P. Perdew, K. Burke and M. Ernzerhof: Phys. Rev. Lett. \textbf{77} (1996) 3865.
\bibitem{VPS} I. Morrison, D. M. Bylander and L. Kleinman: Phys. Rev. B \textbf{47} (1993) 6728.



\bibitem{Gao} G. Gao, K. V. Workum, J. D. S. and J. A. Harrison, J. Phys.: Condens. Matter \textbf{18} (2006) S1737.


\bibitem{G2D_QM1} C. H. Xu, C. Z. Wang, C. T. Chan and K. M. Ho,  J. Phys.: Condens. Matter \textbf{4} (1992) 6047.
\bibitem{G2D_QM2} S. Scandolo, M. Bernasconi, G. L. Chiarotti, P. Focher, and E. Tosatti, Phys. Rev. Lett. \textbf{74} (1995) 4015.
\bibitem{G2D_QM3} Y. Takeyama, T. Ogitsu, K. Kusakabe, and S. Tsuneyuki, Phys. Rev. B\textbf{54} (1996) 14994.
\bibitem{G2D_QM4} A. D. Vita, G. Galli, A. Canning, and R. Car, Nature \textbf{379} (1996) 523.
\bibitem{G2D_QM5} A. D. Vita, G. Galli, A. Canning, and R. Car, Applied Surface Science \textbf{104/105} (1996) 297-303.
\bibitem{G2D_QM6} X. Wang, S. Scandolo, and R. Car, Phys. Rev. Lett. \textbf{95} (2005) 185701.
\bibitem{G2D_QM7} G. Kern and J. Hafner, Phys. Rev. B\textbf{58} (1998) 13167.
\bibitem{G2D_QM8} F. Zipoli, M. Bernasconi, and R. Marton\'{a}k, Eur. Phys. J. B \textbf{39} (2004) 41-47.
\bibitem{G2D_QM9} Yasuaki Omata, Yuichiro Yamagami, Kotaro Tadano, Takashi Miyake, and Susumu Saito, Physica E \textbf{29} (2005) 454-468.




%
%
%
%
%
%
%
%
%
%
%
%


%
\end{thebibliography}
\end{document}